\documentclass{nature}
\usepackage[utf8]{inputenc}
\usepackage{graphicx}
\graphicspath{ {./figures/} }
\usepackage{textcomp,gensymb}
\usepackage[euler]{textgreek}
\usepackage{lineno}

\usepackage{caption}
\DeclareCaptionLabelFormat{vertline}{#1 #2 $|$}
\usepackage{prettyref}
\newrefformat{fig}{Fig.~\ref{#1}}
\newrefformat{exfig}{Extended Data Fig.~\ref{#1}}

\usepackage[normalem]{ulem}

\usepackage{color}

\usepackage{units}
\usepackage{amsmath,amssymb}
\usepackage{comment}

\makeatletter
\let\saved@includegraphics\includegraphics
\AtBeginDocument{\let\includegraphics\saved@includegraphics}
\renewenvironment{figure}{\@float{figure}}{\end@float}

\makeatother

\newcommand{\beginsupplement}{%
        \setcounter{table}{0}
        \renewcommand{\table}{\arabic{table}|}%
        \setcounter{figure}{0}
        \renewcommand{\figurename}{Extended Data Figure}
     }

\title{Tracing out Correlated Chern Insulators in Magic Angle Twisted Bilayer Graphene}

\author{Youngjoon Choi$^{1,2,3*}$, Hyunjin Kim$^{1,2,3*}$, Yang Peng$^{4,3}$, Alex Thomson$^{2,3,5}$, Cyprian Lewandowski$^{2,3,5}$,  
Robert Polski$^{1,2}$, Yiran Zhang$^{1,2,3}$, 
Harpreet Singh Arora$^{1,2}$, Kenji Watanabe$^6$, Takashi Taniguchi$^6$, Jason Alicea$^{2,3,5}$, Stevan Nadj-Perge$^{1,2\dagger}$}

\begin{document}

\maketitle

\begin{affiliations}

    \item T. J. Watson Laboratory of Applied Physics, California Institute of Technology, 
         1200 East California Boulevard, Pasadena, California 91125, USA
    \item Institute for Quantum Information and Matter, California Institute of Technology, Pasadena, California 91125, USA
    \item Department of Physics, California Institute of Technology, Pasadena, California 91125, USA
    \item Department of Physics and Astronomy, California State University, Northridge, California 91330, USA
    \item Walter Burke Institute for Theoretical Physics, California Institute of Technology, Pasadena, California 91125, USA
    \item National Institute for Materials Science, Namiki 1-1, Tsukuba, Ibaraki 305 0044, Japan
    
    \item[*] These authors contributed equally to this work
    \item[$^\dagger$] Correspondence: s.nadj-perge@caltech.edu 
\end{affiliations}


\begin{abstract}
Magic-angle twisted bilayer graphene (MATBG) exhibits a range of correlated phenomena that originate from strong 
electron-electron interactions. These interactions make the Fermi surface highly susceptible to reconstruction 
when $\mathbf{ \pm 1, \pm 2, \pm 3}$ electrons occupy each moir\' e unit cell and lead to the formation of correlated 
insulating, superconducting and ferromagnetic phases\cite{caoCorrelatedInsulatorBehaviour2018, caoUnconventionalSuperconductivityMagicangle2018, 
yankowitzTuningSuperconductivityTwisted2019, luSuperconductorsOrbitalMagnets2019}. While some phases have been shown 
to carry a non-zero Chern 
number\cite{sharpeEmergentFerromagnetismThreequarters2019, serlinIntrinsicQuantizedAnomalous2019}, 
the local microscopic properties and topological character of many other phases remain elusive. 
Here we introduce a set of novel techniques hinging on scanning tunneling microscopy (STM) to map out topological 
phases in MATBG that emerge in finite magnetic field. By following the evolution of the local density of states (LDOS) at
the Fermi level with electrostatic doping and magnetic field, we visualize a local Landau fan diagram that enables us
to directly assign Chern numbers to all observed phases. We uncover the existence of six topological phases emanating from
integer fillings in finite fields and whose origin relates to a cascade of symmetry-breaking transitions driven by 
correlations\cite{zondinerCascadePhaseTransitions2020,wongCascadeTransitionsCorrelated2019}. The spatially resolved 
and electron-density-tuned LDOS maps further reveal that these topological phases can form only in a small range of 
twist angles around the magic-angle value. Both the microscopic origin and extreme sensitivity to twist angle differentiate 
these topological phases from the Landau levels observed near charge neutrality. Moreover, we observe that even the 
charge-neutrality Landau spectrum taken at low fields is considerably modified by interactions and exhibits prominent 
electron-hole asymmetry and an unexpected splitting between zero Landau levels that can be as large as 
$\mathbf{{\sim }\,3-5}$~meV, providing new insights into the structure of flat bands. Our results show how strong electronic 
interactions affect the band structure of MATBG and lead to the formation of correlation-enabled topological phases.

\end{abstract}

When two graphene sheets are rotationally misaligned (twisted), the interlayer coupling
leads to the formation of an effective triangular moir\' e lattice with spatial periodicity $\mathrm{L_m}=a/(2\sin(\theta/2))$ 
set by the twist angle $\theta$  and graphene lattice constant $a=0.246$ nm \cite{lopesdossantosGrapheneBilayerTwist2007,bistritzerMoireBandsTwisted2011}. At small twist 
angles, the moir\' e period is hundreds of times larger than the inter-atomic distance, and the electronic 
bands of the bilayer, by virtue of the moir{\'e} interlayer coupling, are substantially modified. Near the 
magic angle ($\theta_{\mathrm{M}}\approx 1.1\degree$), the electronic structure consists two maximally flat 
bands that give rise to strongly correlated physics and are separated by ${\sim}\, 20-30$ meV gaps from the more 
dispersive remote bands. In addition to modifying the band structure, the periodic moir\' e potential also affects 
the band topology. As shown in experiments where graphene and hexagonal boron nitride (hBN) are aligned\cite{huntMassiveDiracFermions2013, 
ponomarenkoCloningDiracFermions2013, deanHofstadterButterflyFractal2013, wangEvidenceFractionalFractal2015}, the added 
spatial periodicity combined with the orbital motion of electrons in high (${\sim }\,15$ T) magnetic 
fields generates Chern insulating phases characteristic of Hofstadter's spectrum\cite{hofstadterEnergyLevelsWave1976}. 
While similar effects are expected in MATBG\cite{bistritzerMoirButterfliesTwisted2011}, the impact of 
strong correlations on Hofstadter physics and the new phases that may emerge in finite magnetic 
fields is to a large degree unexplored.

\begin{figure}[p]
\captionsetup{format=plain,labelsep=space}
\begin{center}
    \includegraphics[width=16cm]{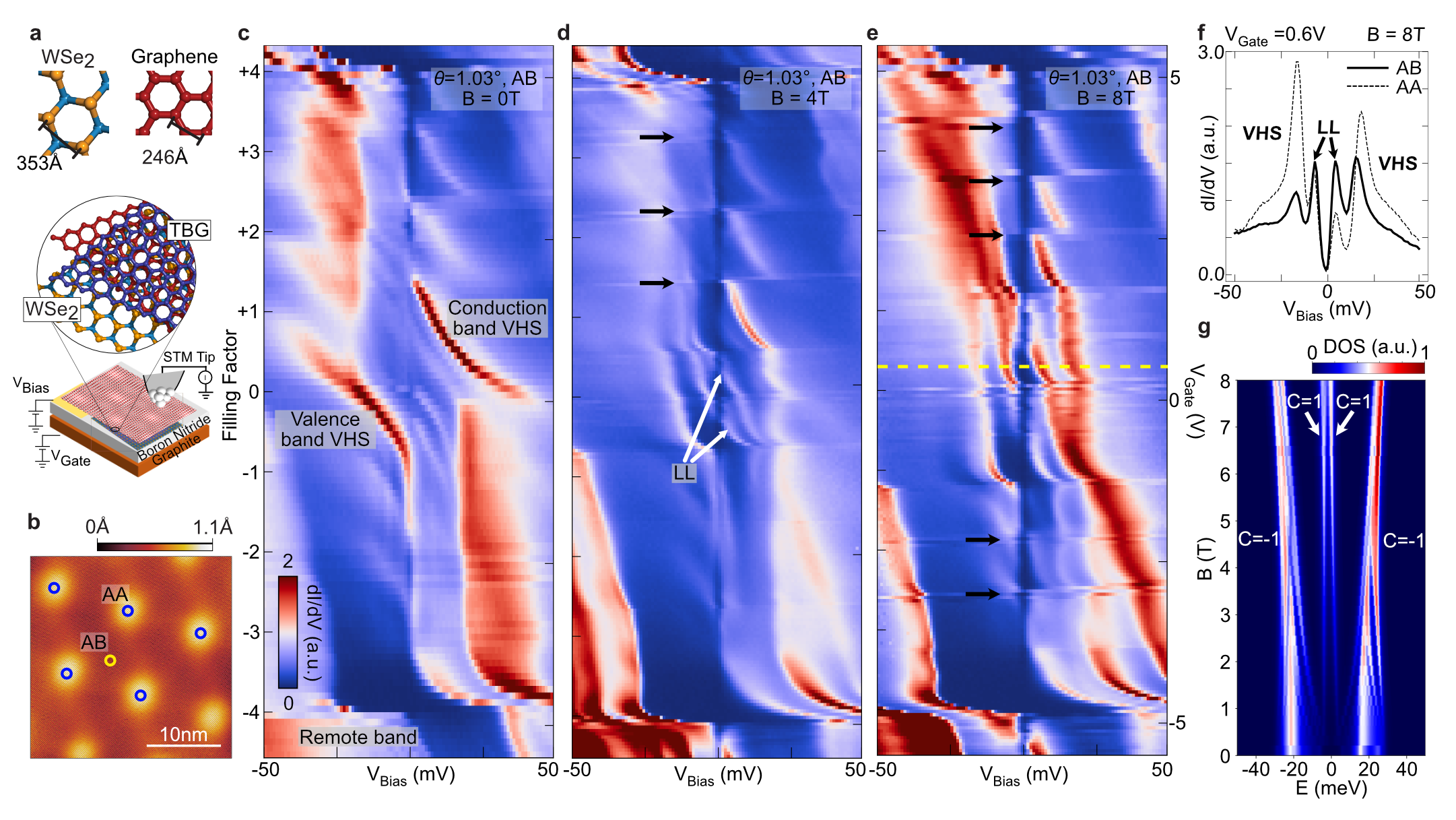}
\end{center}
\caption{{\bf Spectroscopy of MATBG with magnetic field at $2$ K.} {\bf a}, Schematic experimental 
setup. MATBG is placed on a monolayer WSe$_2$ and supported by hBN. A graphite gate resides underneath. Inset shows details of WSe$_2$ and graphene 
crystal structure. {\bf b}, Typical topography showing a moir\'e pattern at the magic angle ($\mathrm{V_{ Bias}} = -400 $ mV, $\mathrm{I} = 20$ pA). 
{\bf c-e}, Point spectra on an AB site at $\theta=1.03\degree$ as a function of $\mathrm{V_{Gate}}$ for
three different magnetic fields applied perpendicular to the sample. 
({\bf c}), $\mathrm{B=0}$ T, the evolution of two peaks in density of states originating from flat-band VHSs. 
As each of the peaks crosses the Fermi energy, it creates a cascade of transitions, appearing here as splitting of the VHSs into multiple branches close to integer filling factors $\nu$. ({\bf d, e}), $\mathrm{B=4}$ T and $\mathrm{B=8}$ T, respectively. 
Landau levels form around charge neutrality ($\nu$ = 0). Black arrows indicate the newly formed gaps that appear 
in magnetic field and are visible as a suppression of $dI/dV$ conductance. 
{\bf f}, Conductance linecuts at $\mathrm{V_{Gate}} = 0.6$ V and $\mathrm{B = 8}$ T on AA and AB sites. 
Landau levels are more visible on the AB site due to reduced VHS weight. {\bf g}, Energy spectrum calculated from the continuum model with parameters chosen such that the relative peak positions match the experimental data in ({\bf f}). Most LLs merge 
into the electron and hole VHSs that each carry Chern number $C=-1$, while two isolated LLs at charge neutrality 
remain around zero energy and carry $C=+1$.}
\label{fig: fig1}
\end{figure}

Figure~\ref{fig: fig1}a shows a schematic of our experiment. 
MATBG is placed on a structure consisting of a
tungsten diselenide (WSe$_2$) monolayer, thick hBN dielectric layer and graphite 
gate (see \prettyref{exfig: fig1} and Methods, section 1, for fabrication details). We use monolayer WSe$_2$ as an 
immediate substrate for MATBG since previous transport studies\cite{aroraSuperconductivityMetallicTwisted2020} 
indicate that WSe$_2$ improves the sample quality and does not change the magic-angle condition. As in previous 
STM studies\cite{brihuegaUnravelingIntrinsicRobust2012,kerelskyMaximizedElectronInteractions2019, choiElectronicCorrelationsTwisted2019}
twist angle can be directly determined by measuring the distance between neighbouring AA sites, where the 
density of states is highly localized, in topographic data (\prettyref{fig: fig1}b). Figure~\ref{fig: fig1}c shows the tunneling 
conductance ($\mathrm{dI/dV}$) corresponding to the local density of states (LDOS) taken at an AB site at zero magnetic field, 
as a function of sample bias ($\mathrm{V_{Bias}}$) and gate voltage ($\mathrm{V_{Gate}}$) that tunes electrostatic 
doping. At $\mathrm{V_{Gate}}>+5$V, the two LDOS peaks originating from the Van Hove singularities (VHS) of the flat 
bands are below the Fermi energy ($\mathrm{E_{F}}$, corresponding to $\mathrm{V_{Bias}} = 0$ mV), indicating 
that flat bands are completely filled with electrons. As $\mathrm{V_{Gate}}$ is reduced, the first VHS corresponding 
to the conduction flat band crosses $\mathrm{E_{F}}$ several times, resetting its position around gate voltages corresponding to occupations of
$\nu = 1, 2, ~\mathrm{and}~ 3$ electrons per moir\' e unit cell (see Methods, section 4, for assigning of filling 
factor $\nu$ to $\mathrm{V_{Gate}}$). 
A similar cascade of transitions was previously observed in STM measurements of MATBG placed directly 
on hBN\cite{wongCascadeTransitionsCorrelated2019}; our observed cascade demonstrates that the addition of WSe$_2$ significantly changes neither the spectrum nor the cascade mechanics. Around the charge neutrality point 
(CNP; $\nu = 0$, corresponding to $\mathrm{V_{Gate}}\approx 0.1$ V) the splitting between two VHS is maximized 
by interactions as discussed previously\cite{choiElectronicCorrelationsTwisted2019,kerelskyMaximizedElectronInteractions2019,
xieSpectroscopicSignaturesManybody2019, jiangChargeOrderBroken2019}. 

When a perpendicular magnetic field is applied, the overall spectrum changes as Landau levels (LLs) develop around 
charge neutrality and $\mathrm{E_F}$. In addition, the onsets of the cascade transitions shift away from the CNP (\prettyref{fig: fig1}d,e, marked by black arrows) 
and are accompanied by a very low LDOS at the corresponding Fermi energies as well as by nearly horizontal resonance peaks, indicating the presence of 
gapped states\cite{jungEvolutionMicroscopicLocalization2011}. We first focus on the LLs. A linecut close to the CNP at $\mathrm{B}=8$~T 
shows four well-resolved peaks (\prettyref{fig: fig1}f); the inner two correspond to LLs. 
The large intensity of VHS peaks on the AA sites obscures some LL-related features (see \prettyref{exfig: fig2} for spectrum on an AA site), so we 
instead study the AB sites to maximize visibility. 

The phenomenological ten-band model\cite{poFaithfulTightbindingModels2019} with parameters chosen to semi-quantitatively match the data 
(\prettyref{fig: fig1}g) suggests that the inner peaks are zero LLs (zLLs) originating from Dirac points while the 
outer peaks (VHS) form from LLs descending from other less-dispersive parts of the band structure and thus can not be individually resolved 
 (see Methods, section 7, for discussion of the modeling). Importantly, both zLLs and VHS are expected to carry 
non-zero Chern number ($+1$ and $-1$ respectively), and to be four-fold spin-valley degenerate. This assignment of the Chern numbers 
takes into account the energetic splitting of the two Dirac cones that is experimentally observed (see discussion on 
\prettyref{fig: fig4}). Note that this splitting naturally accounts for the reduction of an eight- to four-fold 
degeneracy observed in previous transport MATBG experiments. With this interpretation in mind, and as discussed in the remainder 
of the paper, we attribute the shifting of the zero-field cascades (black arrows in \prettyref{fig: fig1}d and e) and the accompanying 
gaps to the formation of Chern insulating phases at high fields, enabled by correlations.

\begin{figure}[p]
\begin{center}
    \includegraphics[width=15cm]{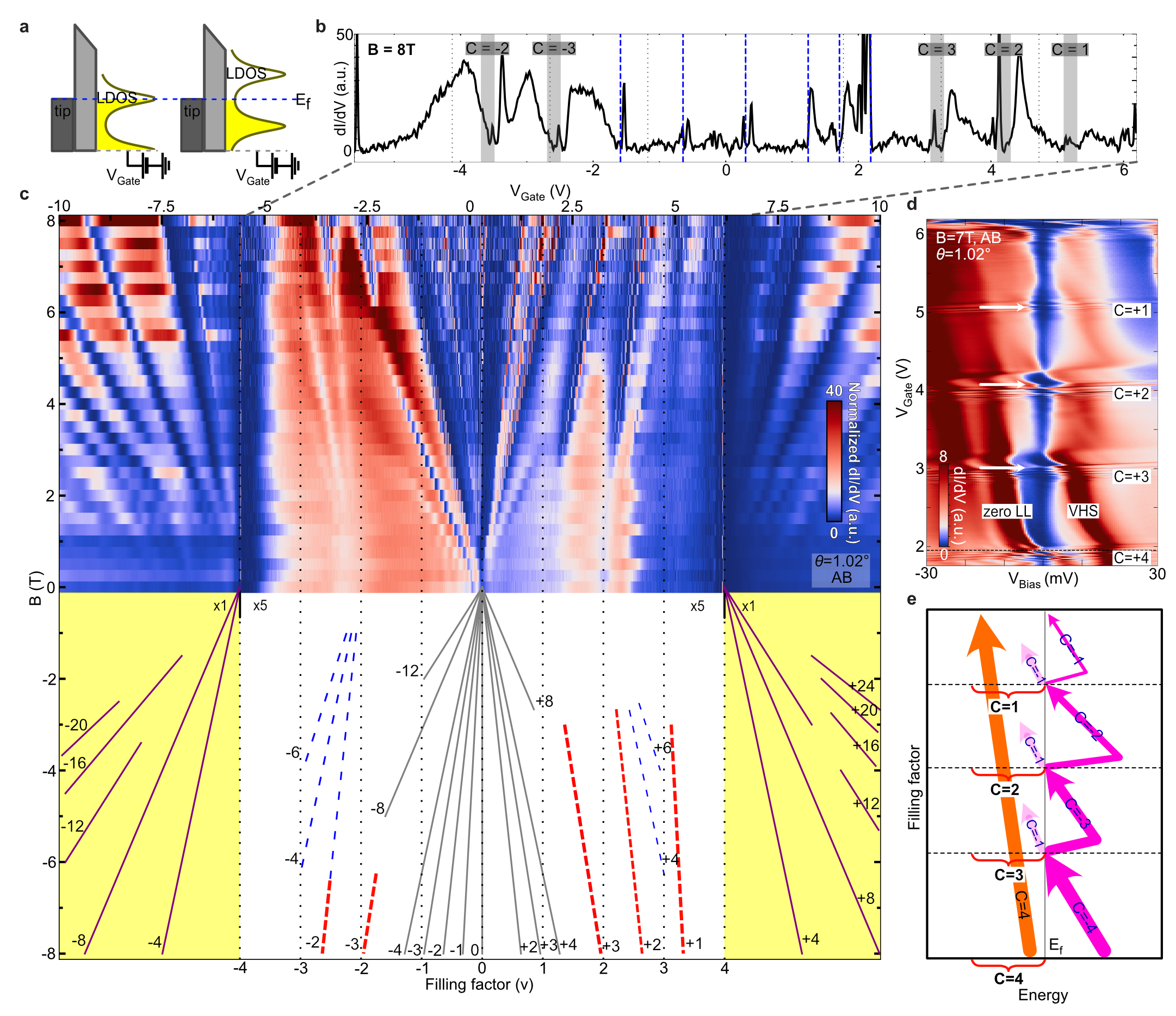}
\end{center}
\caption{{\bf Local Density of States (LDOS) Landau fan and gaps induced by Chern insulating phases.} 
 {\bf a}, Principle of acquiring LDOS Landau fan. Conductance $d\mathrm{I}/d\mathrm{V} \propto$ LDOS is measured while sweeping 
 $\mathrm{V_{Gate}}$ to change carrier density at fixed $\mathrm{V_{Bias}} = 0$ mV.  {\bf b}, Example linecut
 taken at $\mathrm{B = 8}$ T and  -6 V $<\mathrm{V_{Gate}}<6$ V. Insulating phases appear as LDOS dips. 
 Chern insulating phases and LLs are indicated by grey regions and blue vertical lines, respectively. The position of these lines is obtained 
 from the slopes in ({\bf c}). {\bf c}, LDOS Landau Fan diagram at an AB site for $\theta$ = 1.02$\degree$. LDOS data, taken by 
 sweeping $\mathrm{V_{Gate}}$, is normalized by an average LDOS value for each magnetic field (separately for flat and remote bands). Black solid (blue dashed) 
 lines indicate gaps between LLs originating from the CNP (half-filling). Purple lines on the yellow background show the LL gaps in remote (dispersive) bands. 
 Magnetic-field-activated correlated Chern insulator gaps are marked by red dashed lines. 
 The signal in the flat band region is multiplied by five to enhance visibility in relation to the remote bands. {\bf d}, 
 Point spectra on the same AB point as in ({\bf c}) 
 as a function of $\mathrm{V_{Gate}}$ taken at 
 $\mathrm{B = 7}$ T, highlighting the crossing of the electron-side VHS in magnetic field. Chern insulator gaps corresponding 
 to ${C = 3}$, ${C = 2}$ and ${C = 1}$ are indicated by white arrows. The gaps are accompanied by 
 resonances (horizontal features) originating from  quantum dots formed within the insulating bulk (see Methods section 5).  ({\bf e}), 
 Schematic of the cascade in magnetic field. Each time the VHS crosses the Fermi energy, a Chern gap appears 
 and the corresponding band Chern number changes.}
\label{fig: fig2}
\end{figure}

The evolution of LLs with magnetic field in MATBG has so far been studied using transport measurements that can only provide information about 
electronic structure close to the Fermi 
energy\cite{caoCorrelatedInsulatorBehaviour2018, yankowitzTuningSuperconductivityTwisted2019, 
luSuperconductorsOrbitalMagnets2019, uriMappingTwistangleDisorder2020}. As a starting point, we relate these experiments to our STM findings 
by utilizing a novel approach that enables us to measure a full Landau fan diagram via LDOS. This LDOS Landau fan is taken by measuring the tunneling 
conductance without feedback, with tip-sample bias voltage ($\mathrm{V_{Bias}}$) fixed at $0$~mV, such that the STM probes the system at the Fermi energy as it is tuned by changing the 
electron density (though $\mathrm{V_{Gate}}$)\cite{choiElectronicCorrelationsTwisted2019} and magnetic field.
The resulting signal is directly proportional to the LDOS and thus it is suppressed in certain regions of carrier 
densities where gaps develop in the energy spectrum (\prettyref{fig: fig2}a and b). 

Our LDOS Landau fan measurements (\prettyref{fig: fig2}c), taken at one particular AB point on the sample, reproduce many 
features previously established by magneto-transport in 
MATBG\cite{caoCorrelatedInsulatorBehaviour2018,yankowitzTuningSuperconductivityTwisted2019, luSuperconductorsOrbitalMagnets2019, aroraSuperconductivityMetallicTwisted2020} 
despite the fact that here we record a fundamentally different quantity, LDOS, 
instead of the typically measured longitudinal resistance. This approach further enables comparing the value of the twist angle extracted from 
the Landau fan with the corresponding local twist angle seen in topography (agreement is within 0.01$\degree$). However, in addition 
to previously identified Landau levels, we also observe a strong suppression of the LDOS in certain regions, indicating the formation of insulating 
phases emanating from $\nu$ = $\pm 1$, $\pm 2$, and $+3$ that abruptly appear in finite fields  ($\mathrm{B} > 3$~T for $\nu > 0$ 
and $\mathrm{B}> 6$~T for $\nu < 0$; red dashed lines in \prettyref{fig: fig2}c)\cite{saitoHofstadterSubbandFerromagnetism2020, wuChernInsulatorsTopological2020, dasSymmetryBrokenChern2020}. 
The Chern numbers corresponding to these phases, $C = \pm 3$, $\pm 2$, and $+1$, respectively, are assigned directly from the observed slopes using 
the gap positions as a function of $\nu$ and the Diophantine equation\cite{wannierResultNotDependent1978}, 
$\nu(\mathrm{B}) = C\times\mathrm{A_m} \times \mathrm{B}/\phi_0 + \nu(\mathrm{B}=0)$ (see \prettyref{exfig: fig6} for data showing also $C=-1$ state). 
Here $\mathrm{A_m}$ is the moir\'e unit cell area, $\phi_0$ is the flux quantum, and $\nu(\mathrm{B}=0)$ denotes the filling ($\pm 1, \pm 2, + 3$) from which the phases emanate.

To better understand the formation of the observed Chern insulating phases, we now turn to spectroscopic measurements at 
fixed magnetic field, focusing on the conduction flat band (\prettyref{fig: fig2}d). As the gate voltage is increased, 
starting from $\mathrm{V_{Gate}}\approx 2$ V, the four-fold degenerate VHS approaches the Fermi level. Just before 
crossing ($\mathrm{V_{Gate}}\approx 3$ V), a series of small gaps open up accompanied by set of nearly horizontal 
resonance peaks. These resonances are attributed to quantum dot formation in the sample, indicating formation of a 
fully gapped insulating phase around the tip (see Methods, section 5, for more details). As 
we increase gate voltage further, part of the VHS is abruptly pushed up in energy (seen at higher $\mathrm{V_{Bias}}$ 
in \prettyref{fig: fig2}d) reducing its spectral weight as the number of unfilled bands (each featuring one VHS) decreases. 
Similar transitions are observed at around $\mathrm{V_{Gate}}\approx 4$ V and $\mathrm{V_{Gate}}\approx 5.1$ V with the 
spectral weight reducing after each transition. This sequence of transitions is an analogue of the $B=0$ cascade. 
Most importantly however, the onsets of these finite-field cascade transitions are now shifted to new $\mathrm{V_{Gate}}$ 
positions, and hence fillings, that trail the location of nearby Chern insulating phases. This is demonstrated in 
\prettyref{exfig: fig3}, where as magnetic field changes, positions of the Chern insulating phases shift, and the onsets of the
cascade shift accordingly for the conduction band VHS. For the valence band VHS, the onsets of the cascade are hardly affected 
until $\mathrm{B}=6$~T where the Chern insulating phases start to form (see Methods, section 6, for 
additional discussion).

These observations can be explained within the Hofstadter picture described in \prettyref{fig: fig1}f and g, where 
each zero LL and VHS respectively carry total Chern number $C = +4$ and $C = -4$ (the factor 4 reflects spin-valley degeneracy).  
Figure~\ref{fig: fig2}e schematically illustrates the evolution of VHS upon changing $\nu$. When the conduction band VHS is 
empty and all the other bands are filled, the total Chern number of the occupied bands is  $C = +4$ ($-4$ from the valence band 
VHS combined with $+4 \times 2$ from CNP LLs). Consequently, the gap between the LL and VHS will follow the slope 
corresponding to $C = +4$ in the LDOS Landau fan. As $E_F$ increases, all four-fold degenerate conduction bands start to 
become populated equally, until $E_F$ reaches the VHS for the first time. At this point, interactions underlying the 
cascade\cite{zondinerCascadePhaseTransitions2020} shift all carriers to one band only (seen as only one of the four 
bands crossing the Fermi energy) and then the other three bands become unfilled and hence are pushed 
to higher energies. Since the added band carries $C = -1$, the total Chern number is now $C = +3$, and the next 
corresponding gap in the LDOS Landau fan follows an accordingly reduced slope. The sequence is repeated again, creating a 
cascade.

\begin{figure}[p]
\begin{center}
    \includegraphics[width=16cm]{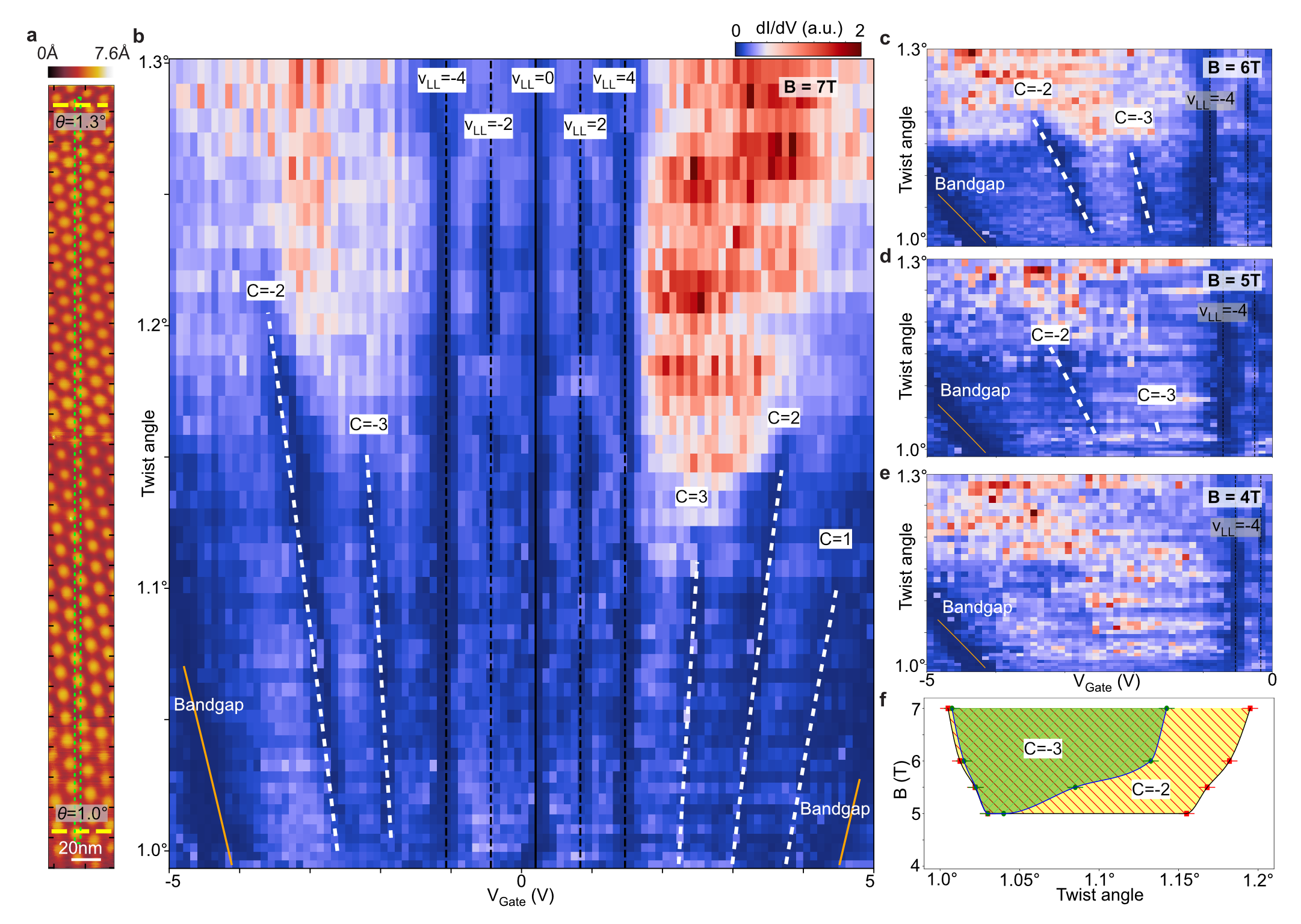}
\end{center}
\caption{{\bf Angle and magnetic-field dependence of the LDOS and identification of Chern insulators.} {\bf a}, 
Topography of a 40nm x 520nm area where twist angle gradually changes from $1.3\degree$ (top yellow line) 
to $0.99\degree$ (bottom yellow line); tunneling conditions are $\mathrm{V_{Bias}} =  100$ mV, $\mathrm{I = 20}$ pA. {\bf b}, Conductance 
at $\mathrm{V_{Bias}}=0$ mV and $\mathrm{B = 7}$ T taken at different spatial points---characterized by different local twist angles $\theta$---and for
different $\mathrm{V_{Gate}}$. Measurements were spatially averaged over the horizontal direction within the green dashed box 
shown in ({\bf a}). Chern insulating gaps develop only in a certain range of angles near the magic angle. The $\mathrm{V_{Gate}}$ positions 
of integer filling factors shift as a function of a local twist angle $\theta$ due to the change of the Chern insulating positions as they depend on the 
moir\'e unit cell area $\mathrm{A_m} = \sqrt{3} L_m^2/2$. Landau 
level gaps originating from the CNP persist for the whole range of angles and do not shift. {\bf c-e}, Magnetic-field dependence of 
the LDOS for hole doping (-5 V $\mathrm{<V_{Gate}<}$ 0 V) at  $\mathrm{B = 6}$ T ({\bf c}),  $\mathrm{B = 5}$ T({\bf d}), and  
$\mathrm{B = 4}$ T ({\bf e}). The Chern insulating gaps disappear as the magnetic field is lowered. {\bf f}, Reconstructed phase 
diagram showing the range of fields and angles where ${C=-2}$ and ${C=-3}$ Chern insulators are observed. In the shaded regions the suppression of 
LDOS due to Chern insulators is prominent. See also \prettyref{exfig: fig6} for high-resolution data resolving the $C=-1$ state.}
\label{fig: fig3}
\end{figure}

\begin{figure}[p]
\begin{center}
    \includegraphics[width=16cm]{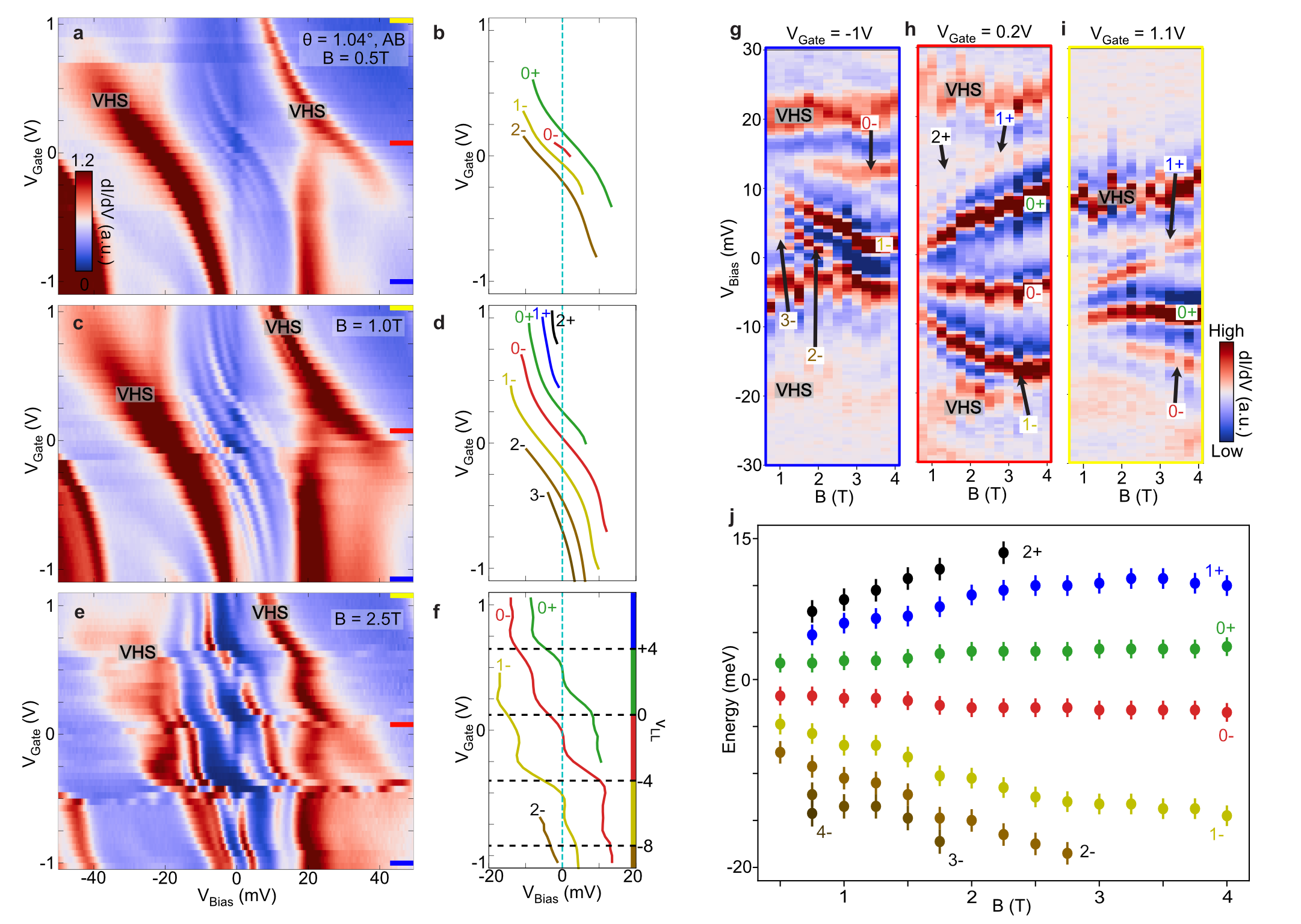}
\end{center}
\caption{{\bf Evolution of Landau levels from charge neutrality}. {\bf a-f}, Point spectra as a function of $\mathrm{V_{Gate}}$ ({\bf a,c,e}) and 
schematics tracking the evolution of Landau levels ({\bf b,d,f}) for $\mathrm{B = 0.5T}$ ({\bf a,b}), $\mathrm{B = 1T}$ ({\bf c,d}), and $\mathrm{B = 2.5T}$ 
({\bf e,f}). At the Fermi energy, each Landau level is equally separated in electron density ($\mathrm{V_{Gate}}$) as marked by the Landau-level 
filling $\nu_{\mathrm{LL}}$, indicating four-fold degeneracy of each level. {\bf g-i}, Linecuts at $\mathrm{V_{Gate}}=-1$ V ({\bf g}), 
$\mathrm{V_{Gate}}=0.2$ V ({\bf h}), and $\mathrm{V_{Gate}}=1.1$ V ({\bf i}) illustrating the LL spectrum change with magnetic field. A smooth 
background was subtracted to enhance visibility. The indicated LLs were identified in ({\bf b,d,f}). {\bf j}, Combined energy spectrum for LLs 
around charge neutrality. Zero energy is set as the midpoint between $0+$ and $0-$ levels at $\mathrm{B=0.5}$ T.}
\label{fig: fig4}
\end{figure}

To verify the role of correlations on the observed Chern insulating phases, we extended the technique introduced in 
\prettyref{fig: fig2}a-c to directly visualize the evolution of the Chern insulating phases with the twist angle. 
In areas where twist angle is slowly evolving over hundreds of 
nanometers (many moir\'e periods), the local angle is well-defined and the strain level is low ($<$0.3$\%$) 
(\prettyref{fig: fig3}a). 
By measuring the LDOS at the Fermi energy against $\mathrm{V_{Gate}}$ and spatial position, we image the evolution of the 
Chern insulating phases as well as LLs from the CNP with twist angle at finite magnetic field (\prettyref{fig: fig3}b). 
Gaps between LLs 
originating from the CNP and corresponding to LL filling factor $\nu_{LL}=\pm 4, \pm 2, 0$ appear at fixed 
$\mathrm{V_{Gate}}$ and 
do not change with the twist angle, as they depend only on electron density. The Chern insulating phases, however, move 
outward 
from the CNP with increasing angle as expected from the change of the moir\'e unit cell size. Also, 
while the gaps between LLs from charge neutrality persist at all angles 
(0.98-1.3$\degree$ in our experiment), the Chern insulating phases are only observed in a certain 
narrow range around the magic angle (\prettyref{fig: fig3}b). For example, the $C = -3$ state emanating from $\nu = -1$ is 
present only for 
$1.02\degree<\theta<1.14\degree$ at $\mathrm{B} = 7$~T while the ${C = -2}$ state is stable for a larger angle range. Moreover, as the 
magnetic field is 
lowered, the angle range where the gaps are observed reduces (\prettyref{fig: fig3}c-e). Figure \ref{fig: fig3}f shows the 
onset in 
field where we observe the ${C = -3}$ and ${C = -2}$ insulating phases, summarizing their angle sensitivity. 

The observed evolution of Chern insulators with twist angle reflects a competition between Coulomb interactions 
and kinetic energy, similarly to the cascade at $B = 0$~T. Here the characteristic scale of electron-electron interactions 
$U$ is approximately set by 
$U \approx e^2/4\pi\epsilon \mathrm{L_m}$ (with $e$, $\epsilon$ being the electron charge and the dielectric
constant, respectively) and increases with increasing twist angle. On the other hand, the typical kinetic energy scale, taken to be the
bandwidth $W$ of the $C=-1$ band that forms the VHS peak seen in measurements, 
shows non-monotonic behavior with twist angle. The bandwidth $W$ is minimal at the magic angle and further narrows with increasing 
magnetic field. We thus expect that magnetic-field-induced Chern insulating phases will occur most prominently close to the magic angle, with larger fields 
required for their onset away from the magic angle as seen in the phase diagram of \prettyref{fig: fig3}f.  
Indeed theoretical estimates of the ratio $U/W$ as a 
function of magnetic field based on a continuum model uphold this reasoning and reproduce the trends observed in the experimental phase 
diagram (see Methods, section 7). Since the correlated Chern insulators occur only when $U/W$ is large, their existence 
serves as an alternative measure of correlation strength. We note that there is a general asymmetry of angle range between the electron 
and hole side (\prettyref{fig: fig3}b). The $C = -2$ phase emanating from $\nu = -2$ on the hole side starts at 1.19$\degree$ and ends 
at $1.01\degree$, while the $C = 2$ state emanating from $\nu = 2$ on the electron side starts at $1.15\degree$ and continues to $0.98\degree$. 
This observation indicates that the ‘magic angle’ condition where correlations are strongest  differs between the conduction and valence 
bands, since the system lacks particle-hole symmetry, and highlights the sensitivity of MATBG physics to tiny twist-angle changes.

Aside from their implications for Chern insulating phases, LLs near charge neutrality observed 
at low magnetic fields ($\mathrm{B}<2$~T) in \prettyref{fig: fig2}c also shed light on the 
MATBG band structure. Some of the early observations in MATBG remain poorly understood---e.g., the appearance of four-fold degenerate LLs around charge 
neutrality\cite{caoCorrelatedInsulatorBehaviour2018} instead of eight-fold as expected from the presence of eight degenerate 
Dirac cones of the two stacked monolayers, and anomalously large bandwidth ($\sim 40$~meV) of the flat 
band\cite{kerelskyMaximizedElectronInteractions2019, choiElectronicCorrelationsTwisted2019, xieSpectroscopicSignaturesManybody2019,
jiangChargeOrderBroken2019} deviating from the $5-10$ meV widths expected from  continuum models\cite{bistritzerMoireBandsTwisted2011}. 
This is largely due to difficulties in band structure calculations that incorporate all relevant effects such as 
electronic correlations\cite{guineaElectrostaticEffectsBand2018,goodwinHartreeTheoryCalculations2020,xieNatureCorrelatedInsulator2020}, 
strain\cite{biDesigningFlatBands2019}, and atomic reconstruction\cite{namLatticeRelaxationEnergy2017}. In particular, several 
mechanisms were proposed to explain the origin of four-fold LLs formed at charge neutrality\cite{hejaziLandauLevelsTwisted2019,zhangLandauLevelDegeneracy2019, 
biDesigningFlatBands2019} but to date there is no general consensus. 

The evolution of LLs from the CNP at $\theta$ = 1.04$\degree$ at low fields appears in \prettyref{fig: fig4}a-f. We name the two LLs 
closest to the CNP as 0+ and 0- (green and red lines in \prettyref{fig: fig4}b,d,f; they develop into the two zero LLs in \prettyref{fig: fig1}f 
at high fields), and then we label the remaining levels sequentially. As already revealed in \prettyref{fig: fig2}c, the four-fold degeneracy 
of each level can also be seen here by the equal separation between LLs in $\mathrm{V_{Gate}}$ at the Fermi level (corresponding $\nu_{LL}$ marked 
in \prettyref{fig: fig4}f), indicating no hidden LLs from poor resolution or layer sensitivity. Having identified LLs, we now obtain the LL energy 
spectrum from linecuts fixing $\mathrm{V_{Gate}}$ at various magnetic fields (\prettyref{fig: fig4}g-i). The relative energy separation 
between LLs changes with B, and different LLs are more visible for different electron densities. When $\mathrm{E_F}$ resides in between 
LLs, the energy separation between those LLs becomes larger due to an exchange interaction (\prettyref{fig: fig4}h, between $0+$ and $0-$). By 
avoiding such interaction-magnified regions and cumulating their relative separations, we  compile the single-particle energy spectrum 
shown in \prettyref{fig: fig4}j (error bars come from small changes in the separation measured at different electron densities). 
Note that for a slightly different strain and angle, fine 
details of the spectrum become different from point to point in the sample, but the energy separations are of similar values.  

The observed LL spectrum is consistent with a scenario wherein the two moir\'e Brillouin zone Dirac cones are shifted in energy, either by strain 
(0.3$\%$ in this area)\cite{zhangLandauLevelDegeneracy2019,biDesigningFlatBands2019} or layer polarization due to a displacement 
field\cite{carrDerivationWannierOrbitals2019a} in our measurements (see Methods, section 9 for more detailed discussion). 
In this scenario, $0+$ and $0-$ (\prettyref{fig: fig4}j) 
come form the two Dirac points. The LL spectrum can be compared to expectations from a Dirac-like 
dispersion $E_n=sgn(n) v_D \sqrt{2e\hbar |n| B} $. The observed energy separations (e.g., $8$~meV for $0-$ to $1-$ and $5$~meV for $1-$ to $2-$ at B$=2$~T) far 
exceed those predicted by a non-interacting continuum model.  
In particular, our measurements yield a Dirac velocity $v_D \approx 1.5-2 \times  10^5 $~m/s that is an order of magnitude larger 
than the continuum-model prediction  (${\sim} 10^4$~m/s), suggesting
strong interactions near charge neutrality. 
Electron-hole asymmetry is also clearly present, as the energy differences between the first few LLs 
on the hole side are comparably larger than their electron-side counterparts. Moreover, upon doping away 
from the CNP, LLs move together toward the VHS while the separation between them is hardly affected (\prettyref{fig: 
fig4}a,c,e). This signals that the shape of dispersive pockets within the flat bands do not change 
significantly as flatter parts of the bands are deformed due to 
interactions\cite{guineaElectrostaticEffectsBand2018,goodwinHartreeTheoryCalculations2020}. 
Taken together, these observations put strict restrictions on the overall band structure of the MATBG and 
provide guidence for further theoretical modeling. Looking ahead, we anticipate that the 
novel STM spectroscopic techniques developed here will enable the exploration of other exotic phases  
in MATBG and related moir\' e systems.

\noindent Note: In the course of preparation of this manuscript we became aware of the related 
work\cite{nuckollsStronglyCorrelatedChern2020}.

\noindent {\bf References:}



\noindent {\bf Acknowledgments:} We acknowledge discussions with Andrea Young, Gil Refael 
and Soudabeh Mashhadi. The device nanofabrication was performed at the Kavli Nanoscience 
Institute (KNI) at Caltech. {\bf Funding:} This work was supported by NSF through grants DMR-2005129 and DMR-1723367 and by the Army 
Research Office under Grant Award W911NF-17-1-0323. Part of the initial STM characterization 
has been supported by CAREER DMR-1753306. Nanofabrication performed by Y.Z. has been 
supported by DOE-QIS program (DE-SC0019166). J.A. and S.N.-P. also acknowledge the support of IQIM 
(an NSF Physics Frontiers Center with support of the Gordon and Betty Moore Foundation through
Grant GBMF1250). A.T., C.L., and J.A. are grateful for support from the Walter Burke 
Institute for Theoretical Physics at Caltech and the Gordon and Betty Moore Foundation’s EPiQS Initiative, 
Grant GBMF8682.  Y.C. and H.K. acknowledge support from the Kwanjeong fellowship.

\noindent {\bf Author Contribution:}  Y.C. and H.K. fabricated 
samples with the help of with the help of R.P., Y.Z., and H.A., and 
performed STM measurements. Y.C., H.K., and S.N.-P. analyzed the data. 
Y.P. and A.T. implemented models. Y.P., A.T., C.L., provided theoretical 
analysis supervised by J.A. . K.W. and T.T. provided materials (hBN). 
S.N-P supervised the project. Y.C, H.K, Y.P., A.T., C.L., J.A. and S.N-P wrote 
the manuscript.

\noindent {\bf Data availability:} The data that support the findings of this 
study are available from the corresponding authors on reasonable request.

\clearpage

\clearpage

\clearpage

\clearpage

\beginsupplement



\begin{figure}[p]
 \captionlistentry{} 
\label{exfig: fig1}
\end{figure}

\begin{figure}[p]
 \captionlistentry{} 
\label{exfig: fig2}
\end{figure}

\begin{figure}[p]
 \captionlistentry{} 
\label{exfig: fig3}
\end{figure}

\begin{figure}[p]
 \captionlistentry{} 
\label{exfig: fig4}
\end{figure}

\begin{figure}[p]
 \captionlistentry{} 
\label{exfig: fig5}
\end{figure}

\begin{figure}[p]
 \captionlistentry{} 
\label{exfig: fig6}
\end{figure}

\begin{figure}[p]
 \captionlistentry{} 
\label{exfig: continuum}
\end{figure}


\begin{figure}[p]
 \captionlistentry{} 
\label{exfig: uoverw}
\end{figure}

\begin{figure}[p]
 \captionlistentry{} 
\label{exfig: scenarios}
\end{figure}

\end{document}